\documentstyle[prl,aps]{revtex}

\def\half{{\textstyle{1\over2}}}
\def\Nhalf{{\textstyle{\frac{N}{2}}}}

\def\onestwo{{\textstyle{1\over\sqrt{2}}}}
\def\beq{\begin{equation}}
\def\eeq{\end{equation}}
\def\Par{{\rm par} \,}

\begin{document}

\title{A Limit on the Speed of Quantum Computation \\ in
Determining Parity\cite{TitleThanks}}

\author{Edward Farhi and Jeffrey
Goldstone\cite{FarhiGoldstoneAuth}}
\address{Center for Theoretical Physics\\ 
Massachusetts Institute of Technology\\
Cambridge, MA  02139}

\author{Sam Gutmann\cite{GutmannAuth}} 

\address{Department of Mathematics\\ 
Northeastern University\\ 
Boston, MA 02115}

\author{Michael Sipser\cite{SipserAuth}}

\address{Department of Mathematics\\ 
Massachusetts Institute of Technology\\
Cambridge, MA  02139}

\maketitle

\begin{abstract}
Consider a function $f$ which is defined on the integers from 1 to $N$ and takes
the values $-1$ and $+1$.  The parity of $f$ is the product over all $x$ from 1 to $N$ of
$f(x)$.  With no further information about $f$, to classically determine the parity of $f$
requires $N$ calls of the function $f$.  We show that any quantum algorithm capable of
determining the parity of $f$ contains at least $N/2$ applications of the unitary operator
which evaluates $f$.  Thus for this problem, quantum computers cannot outperform
classical computers.
\end{abstract}

\pacs{0367.L, 02.70}

\section{Introduction}

If a quantum computer is ever built, it could be used to solve certain problems in less
time than a classical computer.  Simon found a problem that can be solved exponentially faster
by a quantum computer than by the provably best classical algorithm\cite{Simon}.  The Shor
algorithm for factoring on a quantum computer gives an exponential speedup over the best
known classical algorithm\cite{Shor}.  The Grover algorithm gives a speedup for the following
problem\cite{Grover}.  Suppose you are given a function
$f(x)$ with $x$ an integer and $1 \le x \le N$.  Furthermore you know that $f$ is either identically
equal to 1 {\it or\/} it is 1 for $N-1$ of the
$x$'s and equal to
$-1$ at one unknown value of $x$.  The task is to determine which type of $f$ you have. 
Without any additional information about $f$, classically this takes of order $N$ calls of
$f$ whereas the quantum algorithm runs in time of order $\sqrt{N}$.  In fact this $\sqrt{N}$
speedup can be shown to be optimal\cite{BBBv}.

It is of great interest to understand the circumstances under which quantum speedup
is possible.  Recently Ozhigov has shown that there is a situation where a
quantum computer cannot outperform a classical computer\cite{Ozhigov}.  Consider a function
$g(t)$, defined on the integers from 1 to $L$, which takes integer values from 1 to $L$.  We
wish to find the $M^{\rm th}$ iterate of some input, say 1, that is, $g^{[M]} (1)$.  (Here
$g^{[n]} (t) = g (g^{[n-1]} (t))$ and  $g^{[0]} (t) = t$.)  Ozhigov's
result is that if $L$ grows at least as fast as
$M^7$ then any quantum algorithm for evaluating the $M^{\rm th}$ iterate takes of
order $M$ calls of the unitary operator which evaluates $g$; of course the classical
algorithm requires $M$ calls.  Later we will show that our result in fact implies a
stronger version of Ozhigov's with $L=2M$.

In this paper we show that a quantum computer cannot outperform a classical
computer in determining the parity of a function; similar and additional
results are obtained in  
\cite{BealsBCMW} and\cite{BuhrmanCW}. Let
\beq
f(x) = \pm 1 \quad {\rm for} \quad x=1, \dots N \ .
\label{eq:1}
\eeq
Define the parity of $f$ by
\beq
\Par (f) = \prod^N_{x=1} f(x)
\label{eq:2}
\eeq
so that the parity of $f$ can be either $+1$ or $-1$.  The parity of $f$ always
depends on the
value of $f$ at every point in its domain so classically it requires $N$ function calls to
determine the parity.  The Grover problem, as described above, is a special case of the
parity problem where additional restrictions have been placed on the function. 
Although the Grover problem can be solved in time of order $\sqrt{N}$ on a quantum
computer, the parity problem has no comparable quantum speedup.

\section{Preliminaries}

We imagine that the function $f$ whose parity we wish to determine is provided to us
in the form of an ordinary computer program, thought of as an oracle.  We then use a quantum compiler to
convert this to quantum code which gives us the unitary operator
\begin{eqnarray}
U_f |x, +1 \rangle &=& |x, f(x) \rangle \nonumber \\
U_f |x, -1 \rangle &=& |x, -f(x) \rangle \ .
\label{eq:3}
\end{eqnarray}
(Here the second register is a qubit taking the values $\pm 1$.)  Defining
\begin{eqnarray}
 |x, s \rangle &=& \onestwo (|x, +1 \rangle + |x, -1 \rangle) \nonumber \\
\noalign{\hbox{and}}
 |x, a \rangle &=& \onestwo (|x, +1 \rangle - |x, -1 \rangle)\ ,
\label{eq:4}
\end{eqnarray}
we have that
\beq
U_f |x, q \rangle = f(x,q) |x, q \rangle  \qquad q=s,a
\label{eq:5}
\eeq
where
\beq
f(x,s) = 1 \quad {\rm and } \quad f(x,a) = f(x) \ .
\label{eq:6}
\eeq
Therefore in the $|x,q\rangle$ basis, the quantum operator $U_f$ is multiplication by
$f(x,q)$.

Suppose that $N=2$ so that $x$ takes only the values 1 and 2.  Then
\begin{eqnarray}
U_f ( |1, a \rangle + |2,a \rangle) &=& f(1) |1, a \rangle + f(2) |2,a \rangle
\nonumber \\
&=& f(1) (|1, a \rangle + \Par (f) |2,a \rangle) \ .
\label{eq:7}
\end{eqnarray}
Now the states $|1, a \rangle + |2,a \rangle$ and $|1, a \rangle - |2,a \rangle$ are
orthogonal so we see that one application of $U_f$ determines the parity of $f$ although
classically two function calls are required.  See for example\cite{Cemm}.
In section IV this
algorithm is  generalized  for the case of $N$ to determine parity
after $N/2$ applications of $U_f$.  

In writing (\ref{eq:3}) we ignored the work bits used in calculating $f(x)$.  This is
because, quite generally, the work bits can be reset to their $x$ independent
values\cite{Bennett}.  To do this you must first copy $f(x)$ and then run the quantum algorithm
for evaluating
$f(x)$ backwards thereby resetting the work bits.  If this is done then a single
application of $U_f$ can be counted as two calls of $f$.

\section{Main Result}

We imagine that we have a quantum algorithm for determining the parity of a function
$f$.  The Hilbert space we are working in may be much larger than the
$2N$-dimensional space spanned by the vectors $|x,q \rangle$ previously described. 
The algorithm is a sequence of unitary operators which acts on an initial vector
$|\psi_0\rangle$ and produces $|\psi_f\rangle$.  The Hilbert space is divided into two
orthogonal subspaces by a projection operator ${\cal P}$.  After producing
$|\psi_f\rangle$, we measure ${\cal P}$ obtaining either 0, corresponding
to parity $-1$, or 1, corresponding to parity $+1$. 
(Note that $\langle \psi_f | {\cal P}| \psi_f \rangle$ is the probability of 
obtaining 1.)
  We say that the algorithm is
successful if there is an $\epsilon >0$ such that
\begin{eqnarray}
{\rm For~} \Par (f) = +1, \qquad 
\langle \psi_f | {\cal P} | \psi_f \rangle &\ge& \half + \epsilon \nonumber
\\
\noalign{\hbox{and}}
{\rm For~} \Par (g) = -1, \qquad 
\langle \psi_g | {\cal P} | \psi_g \rangle &\le& \half - \epsilon \ .
\label{eq:8}
\end{eqnarray}
This is a \emph{weak} definition of success for an algorithm---we only ask
that the probability of correctly identifying the parity of $f$ be greater
than $\frac{1}{2}$ no matter what $f$ is.  Since we are proving the
\emph{non}existence of a successful  (short) algorithm, our result is
correspondingly strong.

The algorithm is a sequence of unitary operators, some of which are independent of $f$, and
some of which depend on $f$ through the application of a generalization of (\ref{eq:5}). 
We need to generalize (\ref{eq:5}) because we are working in a larger Hilbert space.  In
this larger Hilbert space there are still subspaces associated with $x$ and $q$. (In other
words, there is a basis of the form $|x,q,w\rangle$ where $x=1, \dots N$ and $q=a,s$ and
$w=1, \dots W$ for some $W$, corresponding to the values of the work bits
that the algorithm may use.)  Accordingly there are projection operators
$P_x$ and $P_q$ which obey
$$
P_x^2 = P_x  \ ; \quad P_x P_y = 0 {\rm ~for~} x \ne y \ ; \quad \sum^N_{x=1} P_x = 1
$$
and
\beq
P_q^2 = P_q \ ; \quad P_s P_a = 0  \ ; \quad \sum_{q=s,a} P_q = 1 \ . \nonumber
\eeq
In terms of these projectors we have
\beq
U_f = \sum_x \sum_q f(x,q) P_x P_q 
\label{eq:10}
\eeq
where the sum over $x$ is from 1 to $N$ and the $q$ sum is over $s$ and $a$.  

An algorithm which contains $k$ applications of $U_f$, acting on $|\psi_0\rangle$, produces
\beq
|\psi_f \rangle = V_k U_f V_{k-1} U_f \dots V_1 U_f |\psi_0\rangle
\label{eq:11}
\eeq
where $V_1$ through $V_k$ are unitary operators independent of $f$,
but which may involve the work bits.  For more extensive discussion,
see\cite{Preskill}.

We will now use (\ref{eq:10}) to put
$\langle \psi_f | {\cal P} | \psi_f \rangle$ in a form where we can see
explicitly how it depends on $f$, allowing us to show that (\ref{eq:8})
is impossible if $k$ is too small.  We have
\beq
\langle \psi_f | {\cal P} | \psi_f \rangle = \sum_{x_1 q_1} \sum_{x_2 q_2} \dots
\sum_{x_{2k} q_{2k}}
A(x_1,q_1 \dots x_{2k}, q_{2k}) \prod^{2k}_{i=1} f(x_i, q_i)
\label{eq:12}
\eeq
where
\beq
A(x_1,q_1 \dots x_{2k}, q_{2k}) = \langle \psi_0| P_{x_1} P_{q_1} V_1^{\dagger} \dots V_k^{\dagger} {\cal P} V_k
\dots V_1 P_{x_{2k}} P_{q_{2k}} |\psi_0\rangle \ .
\label{eq:13}
\eeq
Note that $A$ does not depend on $f$.

There are  $2^N$ different possible $f$'s of the form given by (\ref{eq:1}).  We now
sum over all these functions and compute
\beq
\sum_f  \langle \psi_f | {\cal P} | \psi_f \rangle \Par (f) = \sum_f \sum_{x_1 q_1} \dots
\sum_{x_{2k} q_{2k}} A(x_1,q_1 \dots x_{2k}, q_{2k}) \prod^{2k}_{i=1} f(x_i, q_i)
\prod^N_{y=1} f(y) \ .
\label{eq:14}
\eeq
Note that 
\beq
\sum_f f(z) = 0 \quad {\rm for} \quad z=1, \dots N
\label{eq:15}
\eeq
because for each function with $f(z) = +1$ there is a function with $f(z) = -1$.  Similarly if $z_1,
z_2 \dots z_n$ are all distinct, we have
\beq
\sum_f f(z_1) f(z_2) \dots f(z_n)  = 0 \ . 
\label{eq:16}
\eeq
Return to (\ref{eq:14}) and consider the sum on $f$,
\beq
\sum_f  \, \,  \prod^{2k}_{i=1} f(x_i, q_i) \prod^{N}_{y=1} f(y) 
\label{eq:17}
\eeq
where $x_1,x_2 \dots x_{2k}$ and $q_1,q_2 \dots q_{2k}$ are fixed.  For any $i$ with $q_i=s$ we
have $f(x_i,s) = 1$.  Thus (\ref{eq:17})  equals
\beq
\sum_f \, \,  \prod_{i {\rm \ with} \atop q_i=a} f(x_i) \prod^{N}_{y=1} f(y)  \ .
\label{eq:18}
\eeq
Now $f^2(z)=1$ for any $z$ and any $f$.  By (\ref{eq:16}), the sum over $f$ in (\ref{eq:18}) will
give 0 unless each term in the second product can be matched to a term in the first product. 
Since the first product has at most $2k$ terms and the second product has $N$ terms, we see that
if $2k<N$ then the sum over $f$ in (\ref{eq:18}) is 0 and accordingly,
\beq
\sum_f  \langle \psi_f | {\cal P} | \psi_f \rangle \Par (f)  =0\ .
\label{eq:19}
\eeq
This implies that for $2k <N$
\beq
\sum_{f, \Par (f) =+1} \langle \psi_f | {\cal P} | \psi_f \rangle =  
\sum_{f, \Par (f) =-1} \langle \psi_f | {\cal P} | \psi_f \rangle 
\label{eq:20}
\eeq
which means that for $k<N/2$ condition (\ref{eq:8}) cannot be fulfilled.

Equation (\ref{eq:20}) shows that our bound holds even if we further relax
the success criterion given in condition~(\ref{eq:8}).
In any algorithm with fewer than $N/2$ applications of $U_f$, demanding
a probability of success greater than {\it or equal\/} to $1/2$ for every $f$
forces the probability to be $1/2$ for every $f$.

\section{An Optimal Algorithm}

To see that the bound $k<N/2$ is optimal, we now show how to solve the
parity problem with $N/2$ applications of $U_f$.  Here we assume that $N$ is even.  We only
need the states $|x,a\rangle$ given in (\ref{eq:4}) for which
\beq
U_f |x,a\rangle = f(x) |x, a \rangle \ .
\label{eq:new21}
\eeq
Define
\begin{eqnarray}
V|x,a\rangle &=& |x+1, a \rangle \qquad x=1,\dots \Nhalf - 1 \nonumber \\
V|\Nhalf,a\rangle &=& |1, a \rangle \nonumber \\
V|x,a\rangle &=& |x+1, a \rangle \qquad x=\Nhalf+1,\dots N- 1 \nonumber \\
V|N,a\rangle &=& |\Nhalf +1, a \rangle
\label{eq:new22}
\end{eqnarray}
Also let
\beq
|\psi_0\rangle = \frac{1}{\sqrt{N}} \sum^N_{x=1} |x,a\rangle \ .
\label{eq:new23}
\eeq
Now compute $|\psi_f\rangle$ given by (\ref{eq:11}) with $k=N/2$ and for the operators
independent of $f$ take
$$
V_1 = V_2 = \dots = V_{k-1}=V \quad {\rm and} \quad V_k=1 \ .
$$
We then have that
\beq
|\psi_f \rangle = \frac{1}{\sqrt{N}} f(1) f(2) \dots f(\Nhalf) \sum^{N/2}_{x=1} |x,a\rangle +
\frac{1}{\sqrt{N}} f(\Nhalf+1) f(\Nhalf +2) \dots f(N) \sum^N_{x=\Nhalf+1} |x,a\rangle \ .
\label{eq:new24}
\eeq
Therefore if $\Par (f) = +1$, the state $|\psi_f\rangle$ is proportional to $|\psi_0\rangle$ whereas
if $\Par (f) = -1$, then $|\psi_f\rangle$ is orthogonal to $|\psi_0\rangle$.  For the parity projection
operator we take ${\cal P} = |\psi_0\rangle \langle \psi_0|$ and we see that the algorithm
determines the correct parity all the time.  Similarly we can show that if $N$ is odd, then with
$k=(N+1)/2$ applications of $U_f$ we can determine the parity of $f$, but this time we need the
states $|x,s\rangle$ as well as $|x,a\rangle$.

\section{Parity as Iterated Function Evaluation}

Here we are interested in evaluating the $N^{\rm th}$ iterate of a function which maps a set of
size $2N$ to itself.  We show that it is impossible for a quantum computer to solve this
problem with fewer than
$N/2$ applications of the unitary operator corresponding to the function.  As noted above, this is a
considerable strengthening of Ozhigov's result.

We assume an algorithm satisfying the above conditions exists and we obtain a contradiction.  Let
the set of $2N$ elements be $\{ (x,r) \}$ where $x=1, \dots N$ and $r=\pm 1$.  For any $f$ of the
form (\ref{eq:1}) define
\beq
g(x,r) = (x+1, rf(x))
\label{eq:new25}
\eeq
where we interpret $N+1$ as 1.  Note that
\beq
g^{[N]} (1,1) = (1, \Par (f)) \ .
\label{eq:new26}
\eeq
Thus an algorithm which computes the $N^{\rm th}$ iterate of $g$ with fewer than $N/2$
applications of the corresponding unitary operator would in fact solve the parity problem
impossibly fast.

\section{Conclusion}
Grover's result raised the possibility that any problem involving a
function with $N$ inputs could be solved quantum mechanically with only
$\sqrt{N}$ applications of the corresponding operator.
We have shown that this is not the case.  For the parity problem, $N/2$
applications of the quantum operator are required.

\subsection*{Acknowledgment} Three of us are grateful to the fourth.

\end{document}